\documentclass[superscriptaddress,amsmath,amssymb,amsfonts,aps,prc,floatfix,showpacs,number,10pt,3p,final,twocolumn]{elsarticle}

\usepackage{graphics}

\usepackage{amssymb}
\usepackage{amsthm}
\usepackage{upgreek}
\usepackage{float}
\usepackage{amsmath}
\usepackage{subfigure}

\biboptions{sort&compress}

\journal{Nuclear Instruments and Methods A}

\begin{document}

\begin{frontmatter}

\author[label2]{U.~Rizwan}

\title{Characteristics of GRIFFIN high-purity germanium clover detectors}

\author[labelTRI]{A.~B.~Garnsworthy}
\author[label2]{C.~Andreoiu}
\author[labelTRI]{G.~C.~Ball}
\author[label2]{A.~Chester}
\author[label2]{T.~Domingo}
\author[labelGuelph]{R.~Dunlop}
\author[labelTRI]{G.~Hackman}
\author[labelGuelph]{E.T.~Rand}
\author[labelTRI]{J.K.~Smith}
\author[label2]{K.~Starosta\corref{cor1}}
\ead{starosta@sfu.ca}
\cortext[cor1]{Corresponding author: 8888 University Dr., V5A 1S6,
  Burnaby, BC Canada. Tel.:+1-778-782-7789}
\author[labelGuelph]{C.~E.~Svensson}
\author[label3]{P.~Voss}
\author[label2]{J.~Williams}

\address[label2]{Department of Chemistry, Simon Fraser University,
  Burnaby, BC, Canada V5A 1S6} 

\address[labelTRI]{Science Division, TRIUMF, 4004 Wesbrook Mall, Vancouver, B.C., Canada, V6T 2A3}

\address[labelGuelph]{Department of Physics, University of Guelph, Guelph, ON, Canada, N1G 2W1}

\address[label3]{Department of Physics,
  Concordia College, Moorhead, MN, USA 56562}

\begin{abstract}
The Gamma-Ray Infrastructure For Fundamental Investigations of Nuclei, GRIFFIN, is a new experimental facility for radioactive decay studies at the TRIUMF-ISAC laboratory. The performance of the 16 high-purity germanium (HPGe) clover detectors that will make up the GRIFFIN spectrometer is reported. The energy resolution, efficiency, timing resolution, crosstalk and preamplifier properties of each crystal were measured using a combination of analogue and digital data acquisition techniques. The absolute efficiency and add-back factors are determined for the energy range of 80 - 3450~keV. The detectors show excellent performance with an average over all 64 crystals of a FWHM energy resolution of 1.89(6)~keV and relative efficiency with respect to a 3"x3" NaI detector of 41(1)\% at 1.3~MeV.
\end{abstract}


\end{frontmatter}


\section{Introduction}

Large arrays of detectors for gamma-ray measurements coupled with auxiliary particle detection systems provide one of the most powerful and versatile tools for studying exotic nuclei through nuclear spectroscopy at rare-isotope beam facilities.
The Gamma-Ray Infrastructure For Fundamental Investigations of Nuclei, GRIFFIN \cite{Garnsworthy2014}, is a new experimental facility for radioactive decay studies at the TRIUMF-ISAC laboratory \cite{Krucken2014}.

GRIFFIN will be used for decay spectroscopy research with low-energy radioactive ion beams. A high gamma-ray detection efficiency, combined with the unique rare-isotope beams provided by ISAC, will support a broad program of research in the areas of nuclear structure, nuclear astrophysics, and fundamental interactions.

\section{GRIFFIN HPGe clover detector}
\label{sec:clover}

\begin{figure}
\centering
\includegraphics[width=1.0\linewidth]{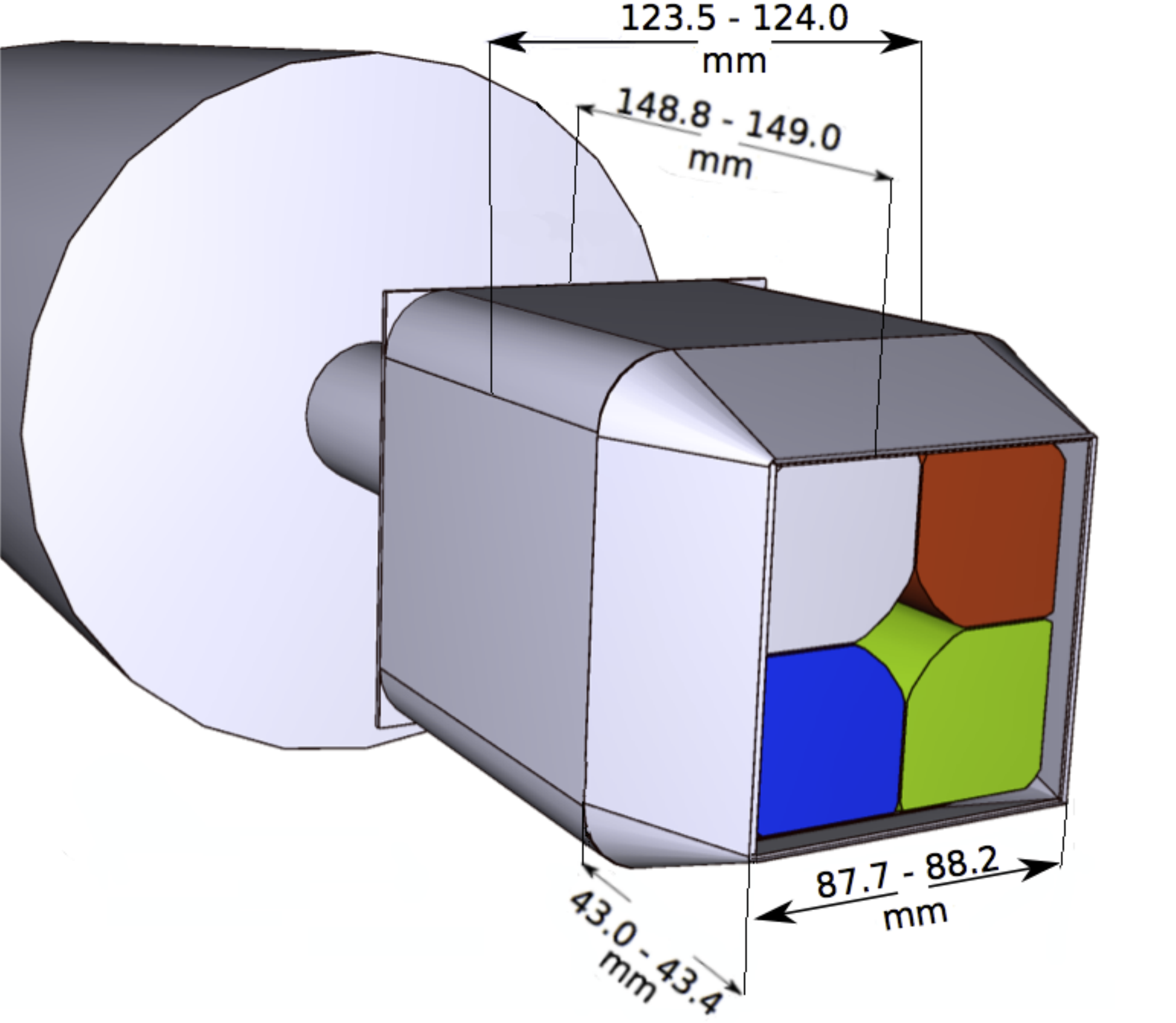}
\caption{3-D model of a GRIFFIN HPGe clover with the exterior dimensional tolerances of the aluminum crystal housing indicated.}
\label{fig:Clover}
\end{figure}

The GRIFFIN detectors were manufactured by Canberra, France-Lingolsheim. The n-type high-purity germanium (HPGe) crystals used in the GRIFFIN detectors are initially 60~mm in diameter and 90~mm in length.
The crystals are machined so that four of them can be housed together in a single cryostat, forming a close-packed ``clover'' arrangement \cite{Duchene99,Shepherd99,Scraggs2005} as shown in Fig.~\ref{fig:Clover}.
The outer edges of the HPGe crystals are tapered at an angle of 22.5 degrees over the first 30~mm of their length to allow for close-packing of neighbouring clover detectors once mounted in the support structure of the GRIFFIN spectrometer. Exterior dimensions of the aluminum crystal housing which contains the vacuum system of the crystals are shown in Fig.~\ref{fig:Clover}. The front face of the aluminum crystal housing has a thickness of 1.5~mm.
The crystals are operated at a temperature of approximately 95~K with cooling provided by a common cold finger from a dewar holding liquid nitrogen. The dewar of each detector has a static holding time in excess of 20~hours.

\begin{figure}
\centering
\includegraphics[width=1.1\linewidth]{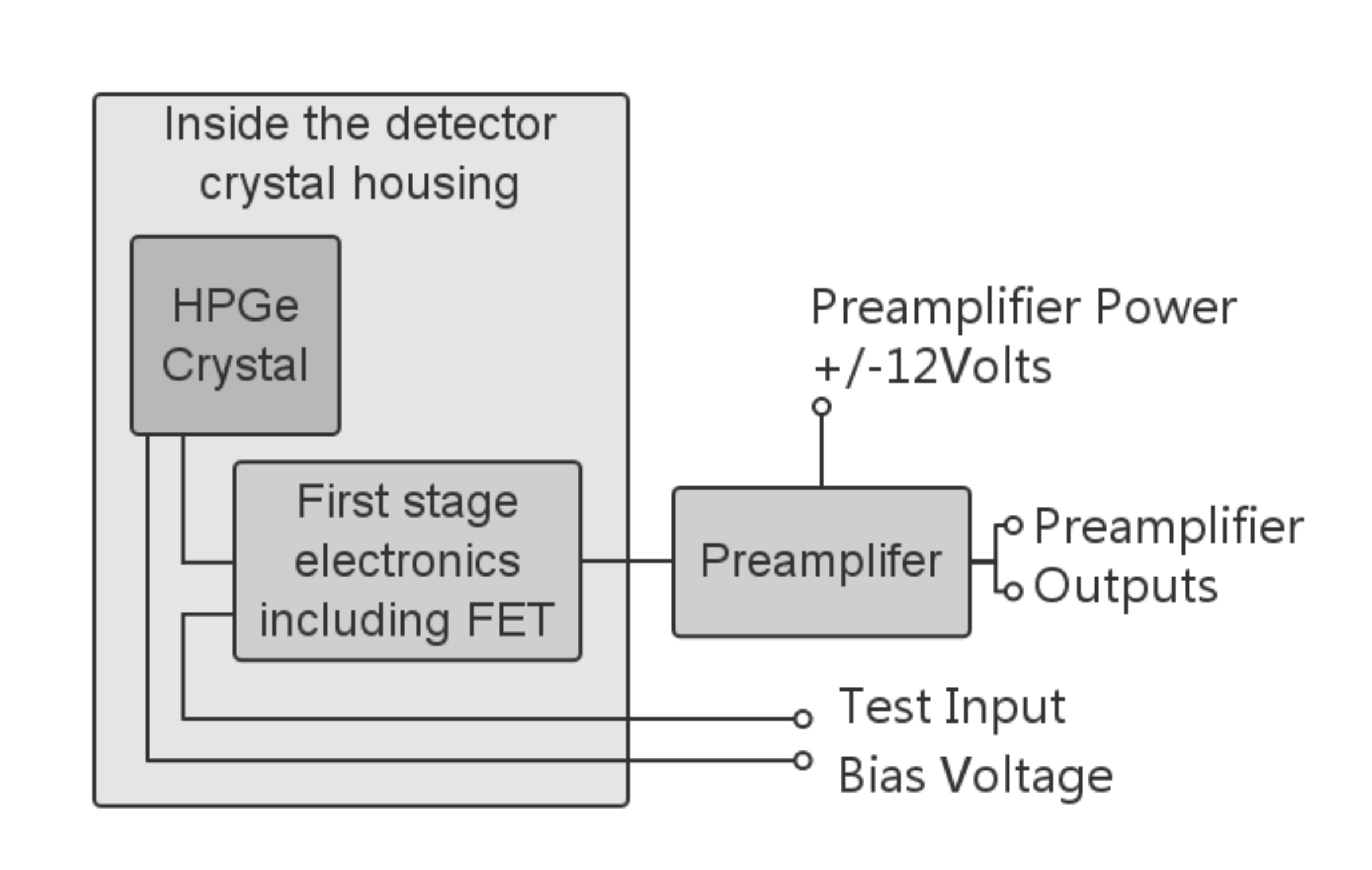}
\caption{Block diagram showing the basic layout of the detector electronics.}
\label{fig:Electronics}
\end{figure}

Fig.~\ref{fig:Electronics} shows a block diagram of the basic layout of the detector electronics. The first stage of the preamplifier electronics is located with the HPGe crystal inside the aluminum crystal housing. The components are in thermal contact with the same cold finger as the crystals and therefore also receive active cooling in order to achieve the best noise conditions of the field-effect transistor. A room-temperature preamplifier is provided for each crystal and located in an enclosure attached to the dewar. No active cooling or heating is provided to this stage of the preamplifier electronics. Each preamplifier has an output signal provided on two bulkhead SMA connectors (which give identical signals) with an output impedance of 50~$\Omega$. An additional SMA bulkhead is provided for each preamplifier to allow injection of a square-wave test pulse signal into the front of the preamplifier as a direct alternative to charge collection from the crystal.

A SHV bulkhead connector is provided for each crystal to allow for individual high voltage bias to be applied to each crystal.
Operating bias voltages are specified by the manufacturer individually for each crystal in the Detector Specification Sheet \cite{Canberra} provided with each detector. The operating bias voltages have values of either +3.5 or +4~kV. During all measurements reported here the bias voltage applied to each crystal was the same as that stated on the Detector Specification Sheet for that crystal.
Each clover is equipped with an internal temperature sensor and associated circuitry which is used for high voltage bias shutdown in the instance of accidental warm up of the detector. Protection from application of bias voltage during a thermal cycle to room temperature is essential as this can lead to electrical discharge that can damage the field-effect transistor.

The four preamplifier chips and bias-shutdown alarm card are located on a common Printed-Circuit Board (PCB) and share a common electrical ground. Preamplifier power at a voltage of $\pm$12~Volts is provided by a 9-pin D-sub connector on the bulkhead.

\section{GRIFFIN clover detector properties}
\label{sec:Acc_tests}

The performance of the GRIFFIN clover detectors was characterized at the Simon Fraser University Nuclear Science Laboratories as part of the initial acceptance testing procedure and is described in this section.

\subsection{Energy resolution}
\label{sec:EnRes}

Energy resolution measurements were performed by placing a series of radioactive sources 25.00(5)~cm from the center of the face of the clover detector.\footnote{The detector manual states that the front of the crystals are 7~mm from this surface.} A $^{152}$Eu source was used for energy resolution measurements at 122.0~keV. A $^{60}$Co source was used for energy resolution determination at 1332.5~keV. In all measurements the total counting rate of each crystal was less than 1~kHz.

\begin{figure}
\centering
\subfigure{\includegraphics[width=0.9\linewidth]{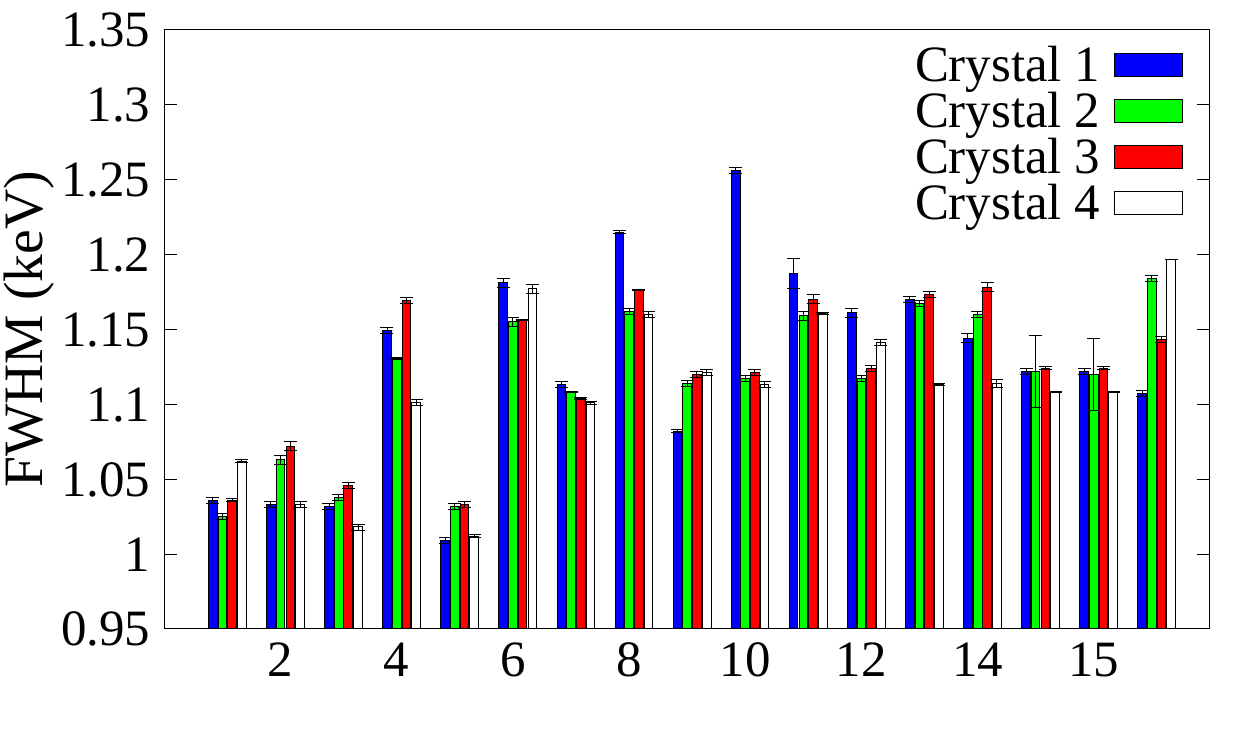}}
\subfigure{\includegraphics[width=0.9\linewidth]{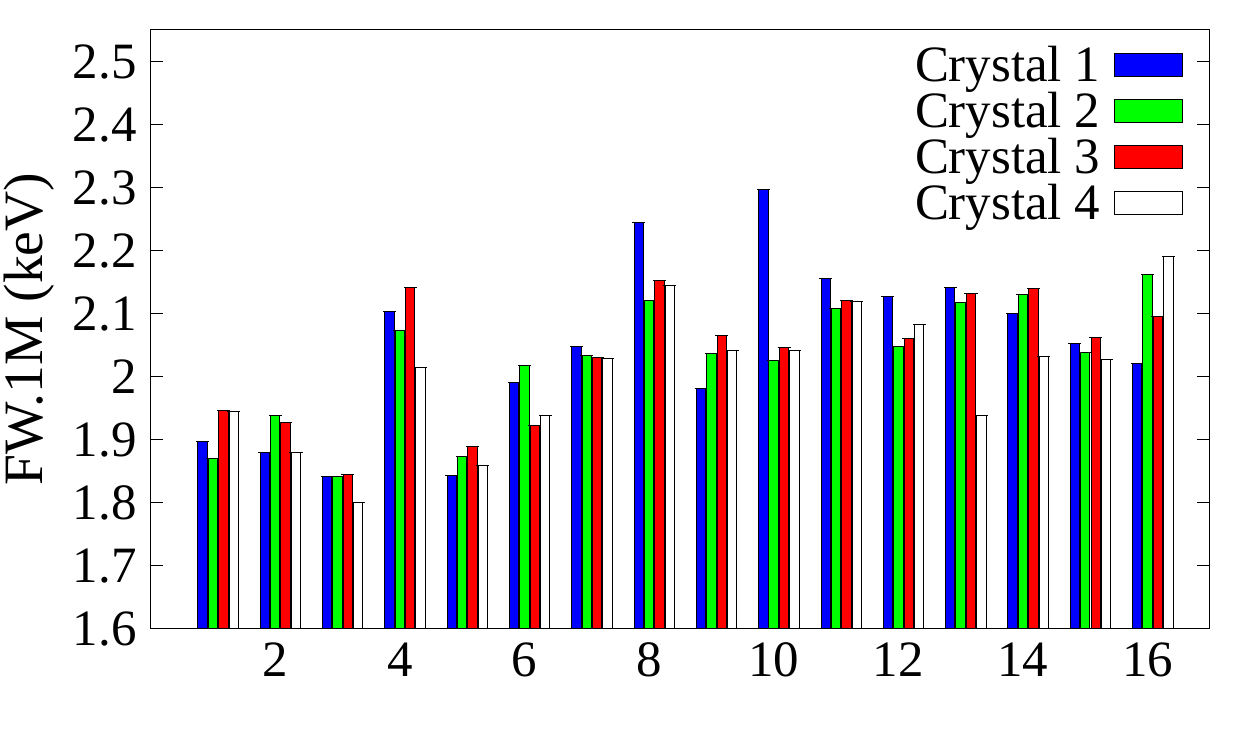}}
\caption{Measured FWHM and FW.1M at 122.0~keV for all crystals. Data were obtained using a $^{152}$Eu source placed 25.00(5)~cm from the center of the face of each clover detector.}
\label{fig:ER122}
\end{figure}

\begin{figure}
\centering
\subfigure{\includegraphics[width=0.9\linewidth]{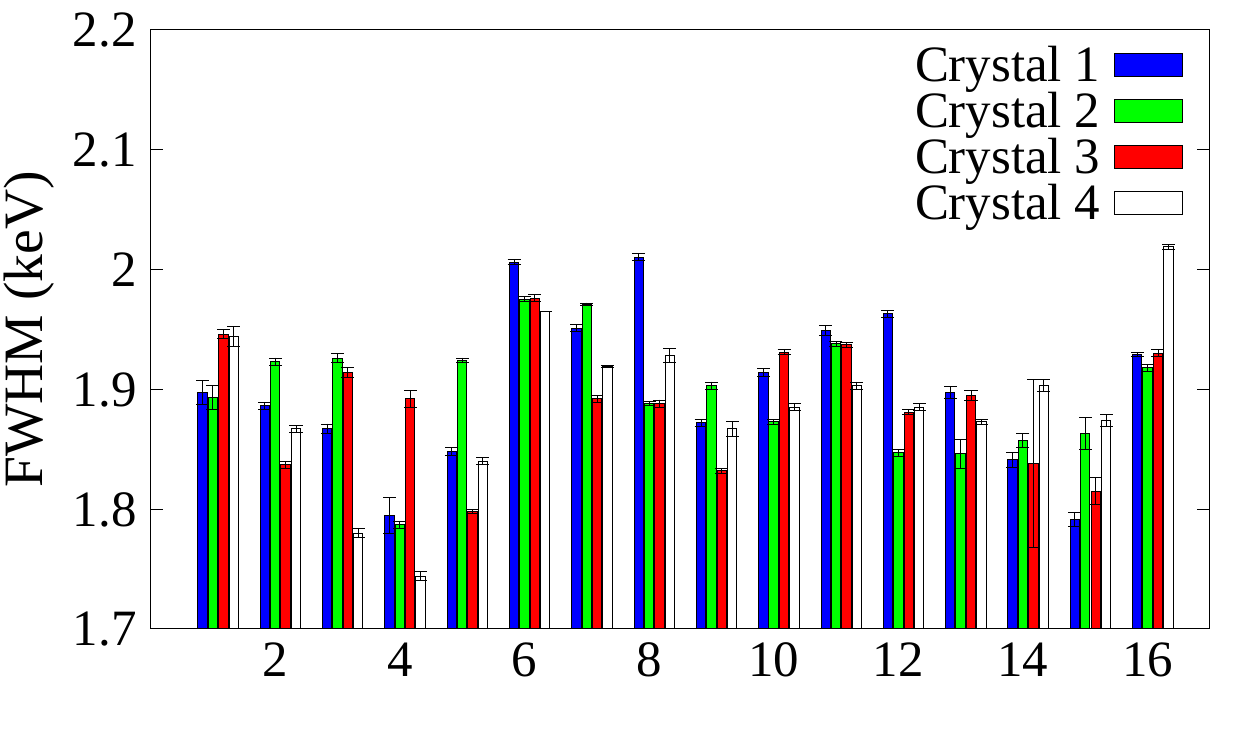}}
\subfigure{\includegraphics[width=0.9\linewidth]{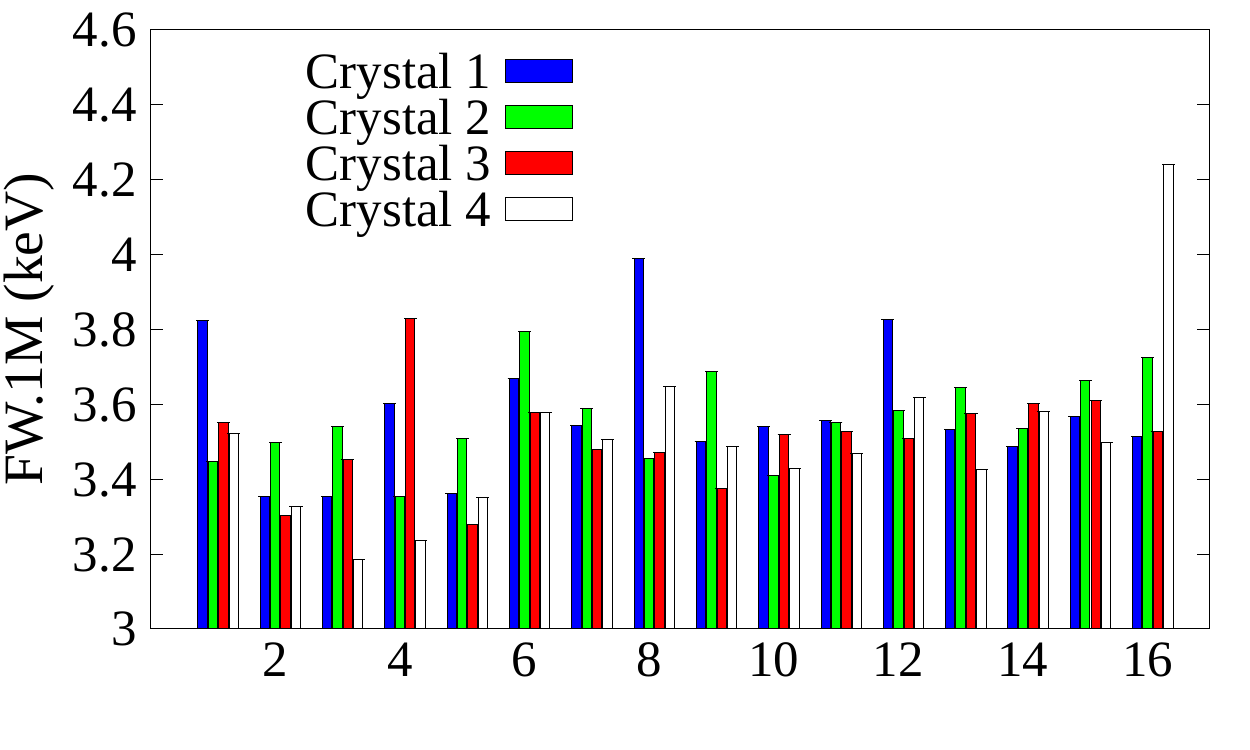}}
\caption{Measured FWHM and FW.1M at 1332.5~keV for all crystals. Data were obtained using a $^{60}$Co source placed 25.00(5)~cm from the center of the face of each clover detector.}
\label{fig:ER1332}
\end{figure}

The preamplifier output signal was fed directly into an ORTEC DSPEC jr 2.0 Multichannel Analyzer (MCA) via a 50~$\Omega$, 5~metre coaxial cable. In separate measurements, a minimum of $10^5$ counts in the background subtracted 122.0~keV and 1332.5~keV photopeaks were collected. The background subtraction was achieved through the automatic background subtraction routines in the Maestro application software \cite{Maestro7}. Energy resolution was defined as the full-width at half-maximum (FWHM) of the photopeak, extracted using gf3, a least square fitting program, from the RADWARE gamma-ray spectroscopy software package \cite{Radford2000}. The full-width at tenth-maximum (FW.1M) resolution was extracted from Maestro's peak fitting algorithm \cite{Maestro7}. The FWHM of the photopeaks for all crystals at 122.0~keV and 1332.5~keV were better than 1.26~keV and 2.02~keV, respectively, while the FW.1M was below 2.3~keV and 4.3~keV, respectively. The results for all crystals are shown in Figs.~\ref{fig:ER122} and \ref{fig:ER1332}. The average energy resolution at 122.0~keV and 1332.5~keV for all 64 crystals is 1.12(6)~keV and 1.89(6)~keV respectively.


\subsection{Efficiency}
\label{sec:EnEff}

\par An activity-calibrated ($\pm$2\% at the 90\% confidence level) $^{60}$Co source was placed in the same source holder used in the energy resolution measurements. Data were collected for each crystal separately without moving the source such that more than $10^5$ counts were accumulated in the 1332.5~keV background-subtracted photopeak. The total counting rate of each crystal was kept below 1~kHz during the data collection.

\begin{figure}
\centering
\includegraphics[width=1.0\linewidth]{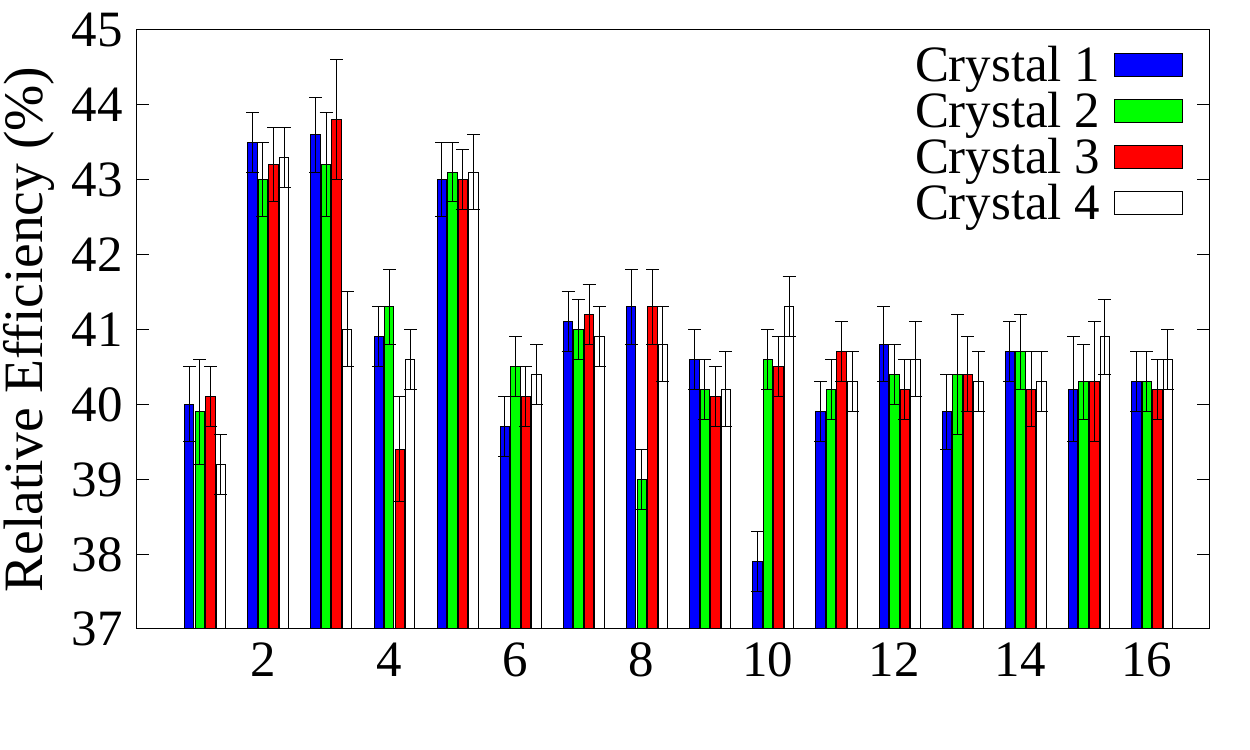}
\caption{Relative efficiency for a 1332.5~keV gamma ray with respect to a 3'' $\times$ 3'' NaI detector. Data were obtained using a calibrated $^{60}$Co source placed 25.00(5)~cm from the center of the face of each clover detector.}
\label{fig:Eff1332}
\end{figure}

The relative efficiency of each GRIFFIN crystal at 1332.5 keV at 25.00(5) cm with respect to a 3" x 3" sodium iodide (NaI) scintillator \cite{IEEE} is presented in Fig.~\ref{fig:Eff1332}. The relative efficiency of all crystals was greater than or equal to 37.9\% at 25.00(5)~cm source-to-detector distance.

\begin{figure}
\centering
\includegraphics[width=1.0\linewidth]{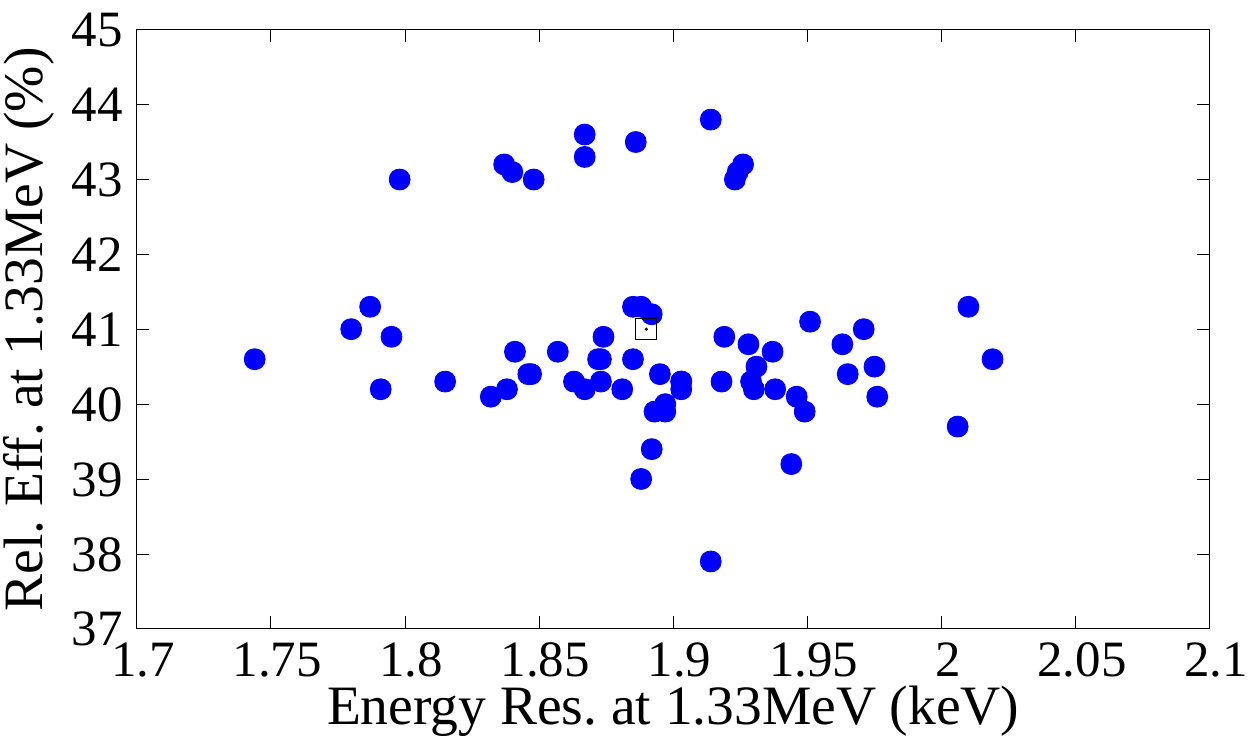}
\caption{Summary of the overall energy resolution and relative efficiency performance at 1.3~MeV of the 64 crystals in the 16 clovers that will form the GRIFFIN spectrometer. The square data point indicates the average value for all 64 crystals.}
\label{fig:Res_vs_Eff}
\end{figure}

Fig. \ref{fig:Res_vs_Eff} summarizes the overall energy resolution and relative efficiency performance at 1.3~MeV of the 64 crystals in the 16 clovers that will form the GRIFFIN spectrometer. The performance of all GRIFFIN detectors is excellent. The average values, indicated by the square data point in Fig. \ref{fig:Res_vs_Eff}, for energy resolution and relative efficiency are 1.89(6)~keV and 41(1)\% at 1.33~MeV respectively.

\subsection{Timing resolution}
\label{sec:Analog_timing_results}

\par
\begin{figure}
\centering
\includegraphics[width=1.0\linewidth]{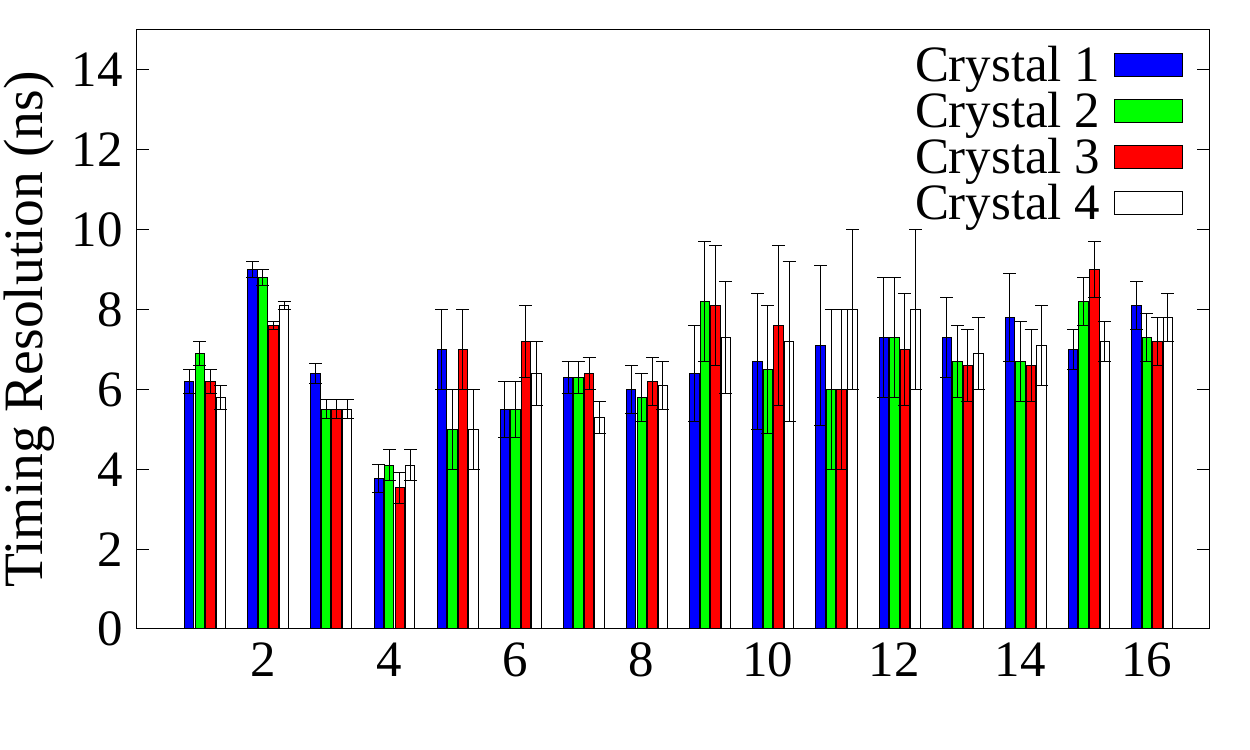}
\caption{Timing resolution of the HPGe crystals for all interactions above 100~keV measured with respect to a BaF$_2$ scintillator. Data obtained using a $^{60}$Co source.}
\label{fig:TimeRes}
\end{figure}

 The timing resolution of each HPGe crystal was measured with respect to a barium fluoride (BaF$_2$) scintillator for all gamma-ray interactions above 100~keV using a $^{60}$Co source. Timing resolution was defined as the FWHM of the HPGe-BaF$_2$ coincidence-timing peak with a minimum of 10$^3$ counts. Standard NIM electronics modules were used for the analogue timing setup in order to obtain the coincidence-timing peak. The timing resolution of each GRIFFIN HPGe crystal with respect to the BaF$_2$ scintillator is given in Fig.~\ref{fig:TimeRes}. The timing resolution of all crystals was below 10~ns for all gamma-ray interactions above 100~keV.

Further investigation of the GRIFFIN clover analogue timing characteristics was performed by applying a stricter coincident energy condition. For one HPGe crystal of each clover, the second preamplifier output signal was used to make a gate on the 1332.5 keV gamma ray. An improvement of 1-3~ns in timing resolution was typically observed when the coincident energy condition was imposed.

\subsection{Inter-Crystal Crosstalk}
\label{sec:cross_talk}

As mentioned in section \ref{sec:clover}, the four preamplifiers share a common electrical ground through a common motherboard PCB and are therefore susceptible to inter-crystal crosstalk. When an interaction occurs in one crystal a baseline offset will be induced on the other three crystals resulting in an apparent energy shift of any signal in those crystals. The magnitude of the energy shift in the second crystal is proportional to the energy deposited in the first crystal and will have a finite time dependance. This effect in the GRIFFIN clovers shows a maximum energy shift of 0.07\% of the deposited energy at a time separation of 5~$\mu$s.
As an example, this corresponds to an energy shift of 0.9~keV for any gamma-ray energy when it is detected 5~$\mu$s after a 1332~keV gamma ray has interacted in one of the other three crystals. This inter-crystal crosstalk will affect the energy resolution in add-back events and at high-counting rates, but it can be accounted for in offline data analysis.

\subsection{Electrical/preamplifier}

A series of electrical and preamplifier properties
were measured as part of the initial acceptance testing procedure for each clover. For all preamplifier properties reported here the uncertainty in the brackets represents the standard deviation of the values measured for all 64 crystals. The uncertainty in the measurement of each property for an individual crystal has a much greater precision.

The peak-to-peak baseline width was determined by observing the output signal of each preamplifier on a 50~$\Omega$ terminated LeCroy LC584AL digital oscilloscope. The peak-to-peak baseline width was less than 3.5~mV for all HPGe crystals with an average value of 2.0(3)~mV.
The DC offset was quantified by observing the shift in baseline for a fully biased-crystal with respect to the ground (zero bias voltage reference). In all cases the DC offset was less than 13~mV with an average absolute offset of 3(4)~mV.

The preamplifier gain was measured by comparing the baseline-to-peak voltage of individual event pulses produced by gamma-ray energies of 662~keV ($^{137}$Cs source) and 1332.5~keV ($^{60}$Co source) measured on a 1~M$\Omega$ terminated oscilloscope. As the output impedance of the preamplifier is 50$\Omega$, the use of a 1~M$\Omega$ termination produces effectively an open circuit for these measurements. The average preamplifier gain of all crystals determined with this method was 208(12)~mV/MeV. Measurements with $^{133}$Ba, $^{152}$Eu and $^{56}$Co sources confirmed the linearity of the gain to better than 0.2~keV over the energy range from 53 to 3451~keV.

The preamplifier output signal rise time, decay-time constant, ringing, and overshoot were examined by connecting a square-wave pulser signal generated by a Hewlett-Packard 8004A Pulse Generator to the test input of each preamplifier. The square-wave signal was generated at a frequency of 1.7~kHz with an amplitude of 80~mV and rise time of 1.20(5)~ns. The preamplifier output signal was digitized using a 14-bit, 100~MHz sampling TIG-10 digitizer in a TIGRESS-style data acquisition system \cite{Martin2007}. The appropriate regions of the digitized waveforms were fitted in an offline analysis to determine the properties of the preamplifiers. The rise time was defined as the length of time for the pulse to increase from 10\% to 90\% of the full amplitude. The decay constant was found from the best fit of an exponential function to the falling part of the waveform. Average values across all 64 preamplifiers of 42(7)~ns and 52(1)~$\mu$s were determined for the rise time and decay constant respectively. The average ringing of the signal (unwanted oscillation following charge collection) was measured to be 0.4(1)\%.
One crystal had an overshoot of 3.8\%, but all other crystals had no measurable overshoot.

\section{Efficiency improvements with add-back}
\label{sec:Add_back}

The dominant interaction process for gamma rays of $\sim$0.2 to $\sim$7~MeV in germanium is by Compton scattering.
In a Compton scattering event it is possible for the gamma ray to deposit part of its energy in one crystal and then enter a neighboring crystal to deposit a portion or the remainder of its energy. These two, or more, energy depositions are time correlated and it is common practice with clover type Ge detectors to recover the full gamma-ray energy by summing the individual events in a procedure known as add-back.

The clover detector can therefore be operated in two modes \cite{Joshi1997},
\begin{itemize}
\item single crystal mode, where the total efficiency of the detector is given by the sum of the individual spectra collected on all four crystals,
\item add-back mode, where total efficiency of the detector is obtained by the sum of reconstructed energies based on observed time correlations.
\end{itemize}
The add-back mode reduces the Compton continuum and the number of counts in the escape peaks while adding counts to the full-energy photopeaks.

A measure of the efficiency enhancement due to the add-back procedure is given by the energy-dependent add-back factor,
\begin{equation}
F(E_\gamma)=\dfrac{N_{\text{AB}}}{N},
\end{equation}
where $E_\gamma$ is the energy of the gamma ray, $N_{\text{AB}}$ is the number of counts in the photopeak after add-back and $N$ is the number of counts in the photopeak before add-back.

Spectra from various sources ($^{133}$Ba, $^{56}$Co, $^{60}$Co and $^{152}$Eu) were used to determine the add-back factors of several GRIFFIN clover detectors over the energy range from 81.0~keV to 3451.2~keV. Data from each source were acquired separately with a 25.00(5)~cm source-to-detector distance. Data from all four HPGe crystals of the GRIFFIN detector were collected simultaneously using a TIGRESS-style data acquisition system \cite{Martin2007} requiring a 500~ns coincident-timing condition. The low-energy threshold for each crystal in these measurements was 26~keV.

\begin{figure} \label{fig:time_peak}
\centering
\includegraphics[width=1.0\linewidth]{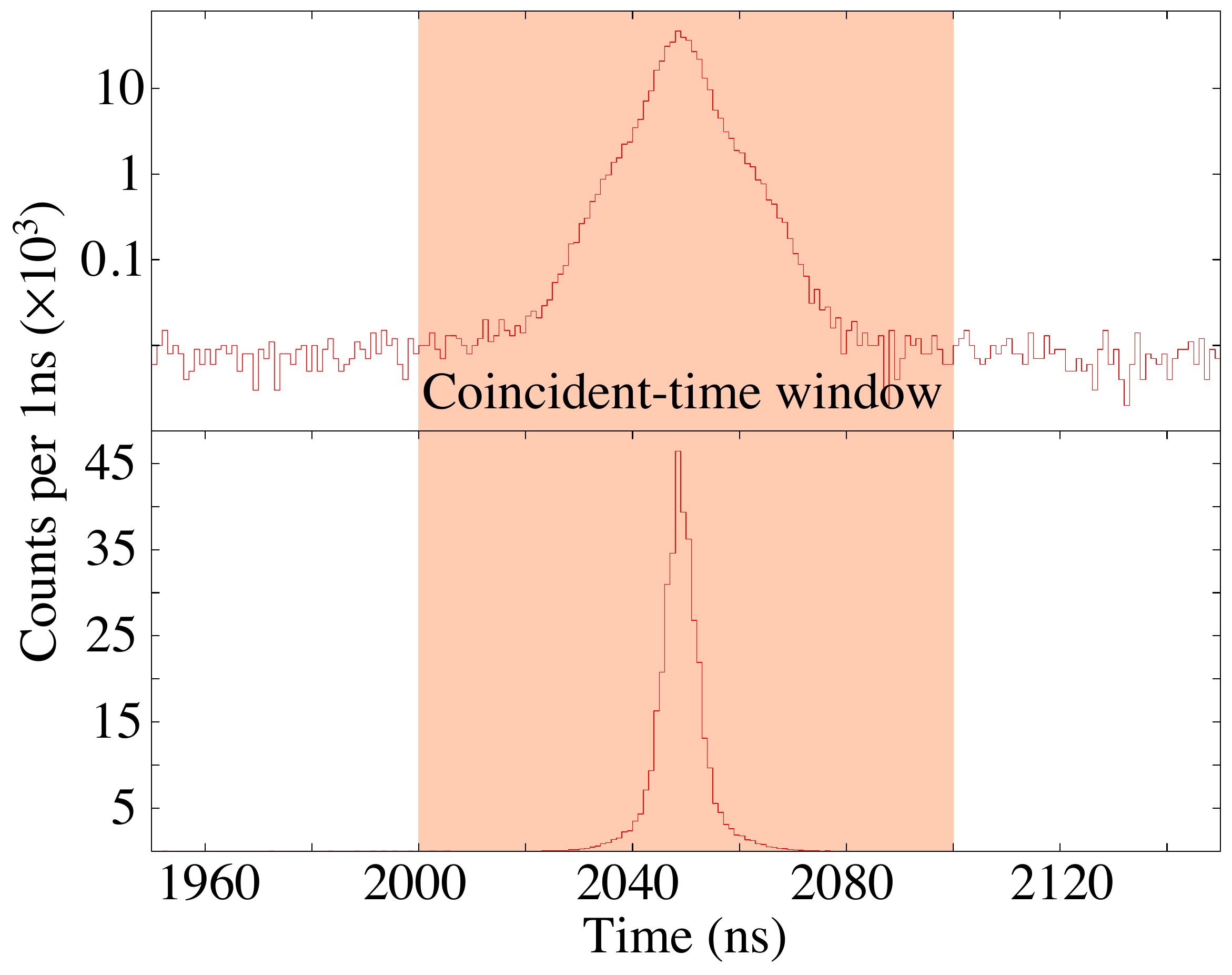}
\caption{HPGe-HPGe coincidence time-peak between Crystal 1 and Crystal 2 of GRIFFIN01. The FWHM is 14.15(6)~ns. The upper and lower panels show the timing-peak with a logarithmic and linear y-axis respectively. The coincident-time window used for add-back was 100~ns.}
\label{fig:time_peak}
\end{figure}

To reduce the contribution from random events detected in the coincident-time window, HPGe-HPGe time-correlations were analyzed. 
Only events that fell within the 100~ns coincident-time window shown in Fig.~\ref{fig:time_peak} were included in the add-back energy reconstruction.

\begin{figure}
\centering
\includegraphics[width=1.0\linewidth]{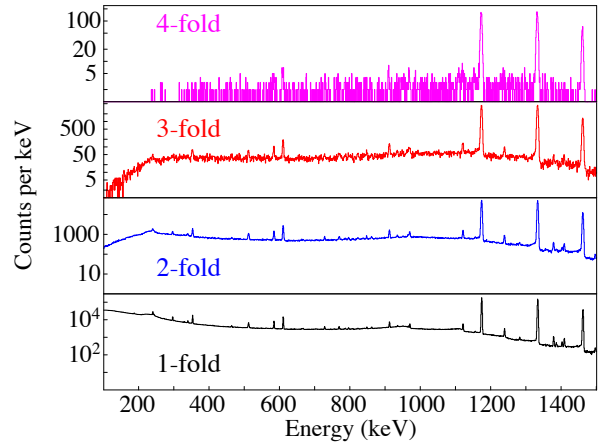}
\caption{$^{60}$Co source spectra of GRIFFIN01 corresponding to 1-fold, 2-fold, 3-fold and 4-fold events. The 1-, 2-, 3- and 4-fold spectra consist of all the time correlated events where one, two, three and four channels produced an event within the coincident-time window, respectively.}
\label{fig:fold_events}
\end{figure}

Fig.~\ref{fig:fold_events} shows example 1-, 2-, 3- and 4-fold spectra for a $^{60}$Co source collected with GRIFFIN detector serial number 1 (GRIFFIN01). The 1-fold spectrum consists of all the events where only one channel detected a hit within the coincident-time window. The 2-, 3- and 4-fold spectra consists of all the time correlated events where two, three or four channels produced an event within the coincident-time window, respectively. An energy offset correction has been applied to the peaks in the $>1$-fold cases as discussed in section \ref{sec:cross_talk}.

\begin{figure}
\centering
\includegraphics[width=1.0\linewidth]{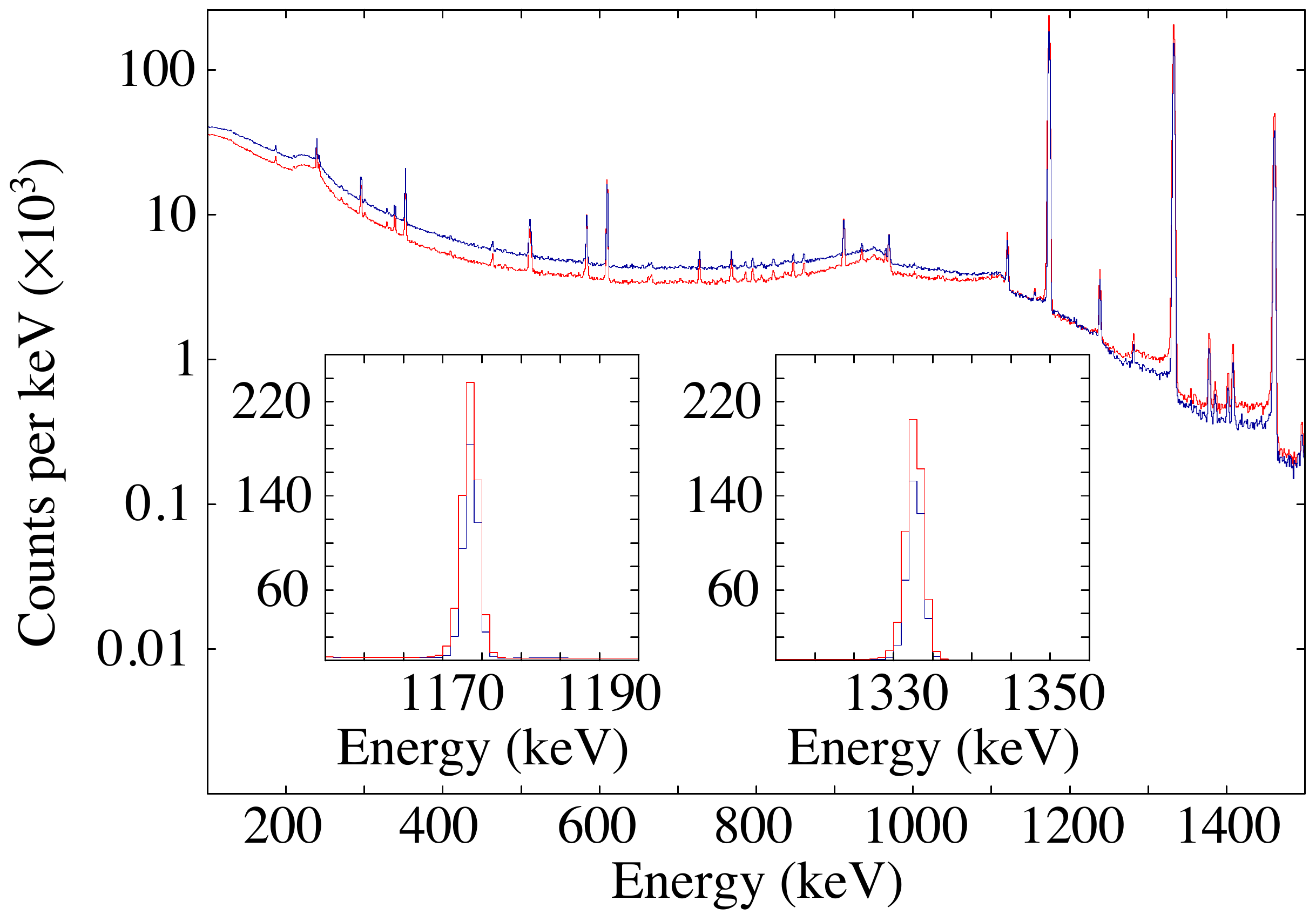}
\caption{(Color online) Spectrum of a $^{60}$Co source collected with GRIFFIN01 operated in singles mode (blue) and in add-back mode (red). Insets show enlargements of the 1173.2~keV and 1332.5~keV photopeaks. The spectrum with add-back is the sum of all the spectra in Fig.~\ref{fig:fold_events}.}
\label{fig:Addback}
\end{figure}

The add-back spectrum is the sum of all $m$-fold spectra ($1 \leq m \leq 4$). Fig.~\ref{fig:Addback} shows the $^{60}$Co spectrum collected on the GRIFFIN01 detector with (red) and without (blue) the add-back procedure performed. A number of gamma rays from room background are also observed in this spectrum. The insets show enlargements of the 1173.2 and 1332.5~keV full-energy photopeaks. It is found that energy resolution worsens by about 0.1-0.4~keV in the add-back mode. It is clear that add-back increases the number of counts in the full-energy photopeaks and reduces the Compton continuum.
At 1332.5~keV, the add-back factor is $\approx$1.4 therefore the efficiency increases by $\approx$40\% at 1332.5~keV when the GRIFFIN clovers are operated in add-back mode instead of singles mode. All GRIFFIN clovers show similar addback factors.

\begin{figure}
\centering
\includegraphics[width=1.0\linewidth]{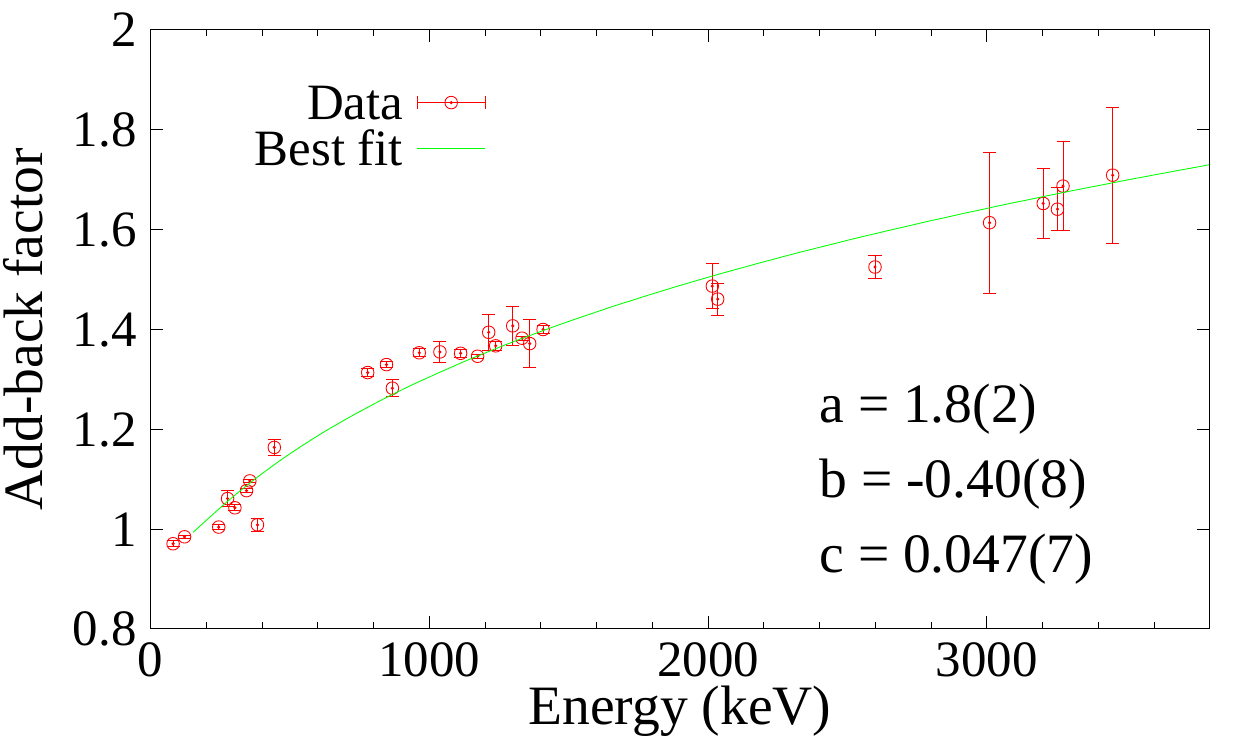}
\caption{Measured add-back factors of the GRIFFIN15 detector. The weighted second-order fit  and best-fit parameters to Eq.~\ref{eq:Add_back_function} are also shown.}
\label{fig:Addback_factor}
\end{figure}

The add-back factors measured over the energy range from 81.0~keV to 3451.2~keV for GRIFFIN15 are shown in Fig.~\ref{fig:Addback_factor}. A weighted second-order function of the form,
\begin{flalign}
\label{eq:Add_back_function}
F &= a + bx + cx^2~~\text{for } E_{\gamma} > 120~\text{keV}, \nonumber \\
x &= \text{ln}(E_{\gamma}),
\end{flalign}
was fitted to the experimental values for gamma-ray energies, $E_{\gamma} >120$~keV, using a least squares fitting routine. The parameters of the best fit were found to be 1.8(2), -0.40(8) and 0.047(7) for $a,b$ and $c$ respectively. These empirical add-back factors are similar to those measured for other clover style HPGe detectors \cite{Joshi1997,SahaSarker2002,Elekes2003,NaraSingh2003}.

\section{Efficiency calibration}
\label{sec:Eff_calib}

\begin{figure}
\centering
\includegraphics[width=1.0\linewidth]{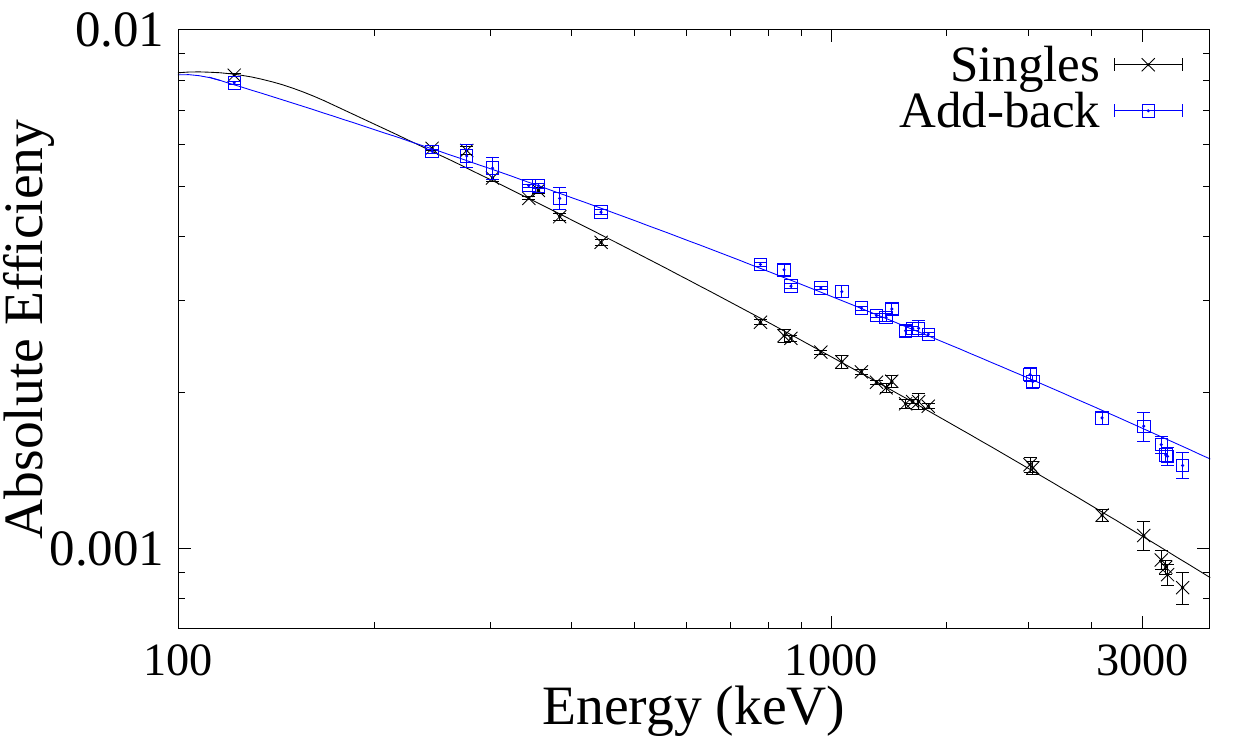}
\caption{Singles (cross) and add-back (open-squares) absolute photopeak efficiency of the GRIFFIN15 clover detector.}
\label{fig:Abs_eff_sin_add}
\end{figure}

Reference~\cite{Usman2015} reports on a method to determine the absolute efficiency curve of a HPGe detector based on a combination of relative efficiency measurements from the $^{133}$Ba, $^{152}$Eu and $^{56}$Co radionuclides and the absolute efficiency measurement from the radionuclide, $^{60}$Co.
This procedure was performed for the GRIFFIN15 detector.
The absolute efficiency was determined with an activity calibrated $^{60}$Co source as described in section \ref{sec:EnEff}.
Further measurements were conducted using uncalibrated sources of $^{133}$Ba, $^{152}$Eu and $^{56}$Co to cover the energy range from 53.2~keV to 3451.2~keV. The activity of all source was $\approx$1--2~$\upmu$Ci (37-74~kBq).

The efficiency data were collected separately for each source with a source-to-detector distance of 25.00(5)~cm. This distance was short enough to allow sufficient statistics to be acquired in a reasonable amount of time while minimizing summing effects. Summing events are the result of two or more gamma rays emitted in the same decay being detected in a single crystal. The effect of summing is to reduce the number of counts in the full-energy photopeak and thus distort the absolute efficiency measurement. In the setup used here the summing corrections are comparable to the statistical errors and smaller than the error of the activity calibration of the source.

Fig.~\ref{fig:Abs_eff_sin_add} shows the absolute efficiency response of the GRIFFIN15 detector when operated in singles mode (stars) and in add-back mode (open squares). At gamma-ray energies around 120~keV, the photopeak efficiency in singles and add-back modes are identical within the precision of the measurements. This means that the add-back factor $F$ is equal to 1 around 120~keV, consistent with the results presented in Section~\ref{sec:Add_back}. It is clear that operating in add-back mode increases the photopeak efficiency, especially at higher energies, and is a beneficial feature of the clover type HPGe detector.

\section{Conclusion}

Sixteen GRIFFIN clover detectors underwent initial acceptance testing at the Simon Fraser University Nuclear Science Laboratories over the 2012-2014 time period. All 16 detectors showed excellent performance and were accepted before the end of 2014. The detectors are now operating in the GRIFFIN spectrometer \cite{Garnsworthy2014} for radioactive decay studies of stopped rare-isotope beams produced by the TRIUMF-ISAC facility.

The performance characteristics of the crystals and preamplifiers are reported. The detectors showed excellent performance with an average over all 64 crystals of a FWHM energy resolution of 1.89(6)~keV and relative efficiency with respect to a 3"x3" NaI detector of 41(1)\% at 1.33~MeV.

\section{Acknowledgements}
 The authors would like to thank Canberra, France-Lingolsheim for the excellent quality and timely manufacture of the GRIFFIN HPGe clover detectors.
This work was supported by the Natural Sciences and Engineering Research Council of Canada (NSERC). The GRIFFIN infrastructure has been funded jointly by the Canada Foundation for Innovation (CFI), TRIUMF and the University of Guelph. TRIUMF receives funding via a contribution agreement through the National Research Council (NRC) of Canada.




\bibliographystyle{elsarticle-num}
\bibliography{GRIFFIN_HPGe_NIM}







\end{document}